# PHASE SPACE FORMULATION OF FILTERING. INSIGHT INTO THE WAVE-PARTICLE DUALITY

D. Dragoman* – Univ. Bucharest, Physics Dept., P.O. Box MG-11, 76900 Bucharest, Romania

**Abstract**:

A phase space formulation of the filtering process upon an incident quantum state is developed. This formulation can explain the results of both quantum interference and delayed-choice experiments without making use of the controversial wave-particle duality. Quantum particles are seen as localized and indivisible concentrations of energy and/or mass, their probability amplitude in phase space being described by the Wigner distribution function. The "wave" or "particle" nature appears in experiments in which the interference term of the Wigner distribution function is present or absent, respectively, the filtering devices that modify the quantum wavefunction throughout the set-up, from its generation to its final detection, being responsible for the modification of the Wigner distribution function.

* Correspondence address: Prof. D. Dragoman, P.O. Box 1-480, 70700 Bucharest, Romania, email: danieladragoman@yahoo.com

## 1. Introduction

The phase space formulation of quantum mechanics, although officially recognized as one of the many (at least nine) formulations of the quantum theory (see [1] and the references therein), has received considerable less attention than the equivalent and much better known Schrödinger or Heisenberg formulations, being mainly used to study the quantum-classical correspondence [2] or dissipative quantum systems [3]. The reason is that this formulation of quantum mechanics is expressed in terms of quasi-probability distributions defined on the classical phase space, namely the phase space spanned by the classical coordinate vector $\boldsymbol{q}$ and the momentum vector $\boldsymbol{p}$. The phase space formulation of quantum mechanics has at first sight little to do with standard quantum mechanics expressed in terms of vectors on the Hilbert space, but, on the other hand, is especially suited to deal with problems connected to the classical limit of the quantum theory.

Although there is no unique phase space quasi-probability distribution, the best known and simpler in many respects is the Wigner distribution function (WDF) [4], which corresponds to the Weyl or symmetric ordering of the non-commuting $\hat{\boldsymbol{q}}$ and $\hat{\boldsymbol{p}}$ quantum operators. This paper demonstrates that the WDF can be employed to formulate in phase space the process of filtering an incoming quantum state and, in particular, to solve one of the "mysteries" of quantum mechanics and quantum optics, i.e. quantum interference [5]. The demonstration is based on the fact that the possibility of interference in either the spatial or momentum domains is indicated by the presence of the interference term in the WDF of a superposition of quantum states. This interference term is filtered away by any device that prevents the passage of one of the interfering states although the transmission of such a filtering device does not vanish over the phase space area on which the interference term is located. The wave or particle behavior of photons or quantum particles resides thus not in different manifestations of the quantum state but in the way in which filtering devices affect or not the interference term in the WDF; the quantum particles or the photons are seen as localized and indivisible concentrations of mass and/or energy that have physical reality and that are distributed in phase space with a probability amplitude given by the WDF. The particular role of the filtering devices in the outcome of measurements that reveal either the "wave" or the "particle" nature of quantum states is best emphasized in delayed-choice experiments. The phase space formulation of quantum mechanics can also explain the result of these experiments, confirming the conclusion already drawn in Ref.6 that quantum interference can best be described in phase space. The paper employs the WDF as the quasi-probability distribution in phase space and extends the results of the more sketchy work in Ref.7, which treats quantum interference as a phase space filtering process.

## 2. Mathematical formalism

The WDF of a quantum system with $n$ degrees of freedom characterized by a wavefunction $\Psi(\boldsymbol{q};t) = \Psi(q_1, q_2, ..., q_n; t)$ is defined as [4]

$$W(\boldsymbol{q},\boldsymbol{p};t) = \frac{1}{h^n} \int \Psi^*\left(\boldsymbol{q}-\frac{\boldsymbol{x}}{2};t\right) \Psi\left(\boldsymbol{q}+\frac{\boldsymbol{x}}{2};t\right) \exp(-i\boldsymbol{p}\boldsymbol{x}/\hbar) d\boldsymbol{x}$$

$$= \frac{1}{h^n} \int \Phi^*\left(\boldsymbol{p}-\frac{\boldsymbol{r}}{2};t\right) \Phi\left(\boldsymbol{p}+\frac{\boldsymbol{r}}{2};t\right) \exp(i\boldsymbol{q}\boldsymbol{r}/\hbar) d\boldsymbol{r} . \qquad (1)$$

where

$$\Phi(\boldsymbol{p}) = h^{-n/2} \int \Psi(\boldsymbol{q}) \exp(-i\boldsymbol{p}\boldsymbol{q}/\hbar) \qquad (2)$$

is the Fourier transform of the quantum wavefunction and $\boldsymbol{qp}$ is a shorthand notation for $q_1 p_1 + q_2 p_2 + ... + q_n p_n$. The WDF can be defined also for mixed states characterized by a Hermitian density operator $\hat{\rho}$ as

$$W(\bm{q},\bm{p};t) = \frac{1}{h^n}\int\left\langle \bm{q}+\frac{\bm{x}}{2}\middle|\hat{\rho}(t)\middle|\bm{q}-\frac{\bm{x}}{2}\right\rangle \exp(-i\bm{p}\bm{x}/\hbar)d\bm{x} . \tag{3}$$

The WDF, as function of the classical phase space coordinates, carries the same information as the quantum state description through a vector on the Hilbert space, being exactly analogous to the density matrix of Dirac and von Neumann [8]. A formulation of quantum mechanics alternative to that of Heisenberg and Schrödinger can therefore be developed in terms of the WDF; the properties and applications of the WDF as well as its relations to other phase space distribution functions are reviewed, for example, in [9-12]. Although analogous to Gibbs phase space density [13], the real-valued quantum-mechanical WDF is not a true density in phase space since it is not positive defined. Its eventual negative values have been interpreted either as suggesting the impossibility of simultaneously measuring conjugate variables such as position and momentum [8] or as indicating the occurrence of phase space interference between different minimum phase space areas covered by a given state [14]. (The Heisenberg uncertainty relation imposes that a quantum state (or its WDF) cannot occupy a phase space region smaller than a quantum blob [15]).

The wavefunction can be recovered from the WDF as

$$\Psi(\bm{q})\Psi^*(\bm{0}) = \int W(\bm{q}/2,\bm{p})\exp(i\bm{p}\bm{q}/\hbar)d\bm{q} , \tag{4}$$

where $\bm{0} = (0,0,...,0)$, a similar relation existing for its Fourier transform $\Phi(\bm{p})$. Moreover, since the squared modulus of the wavefunction and its Fourier transform, interpreted as probability distributions in the position and momentum domains, respectively, are expressed in terms of the WDF through

$$|\Psi(\bm{q})|^2 = \int W(\bm{q},\bm{p})d\bm{p} , \qquad |\Phi(\bm{p})|^2 = \int W(\bm{q},\bm{p})d\bm{q} , \tag{5}$$

it follows that the WDF can be regarded as the quasi-probability (or probability amplitude) of finding a quantum particle in phase space. The normalization condition

$$\int W(\boldsymbol{q},\boldsymbol{p})d\boldsymbol{q}d\boldsymbol{p} = 1 \tag{6}$$

strengthens this interpretation, valid (according to the author's viewpoint) for either individual quantum particles or ensembles of quantum particles. The possible negative values of the WDF are not at odds with this interpretation since not the WDF itself but averages of it over phase space regions indicate the probability of finding a quantum particle in that region; these averages are always positive if taken over phase space regions larger than or equal to a quantum blob. (A quantum blob is any admissible subset of the phase space region with a projection area on any of the conjugate planes $q_j$, $p_j$ equal to $h/2$. In particular, for $n = 1$ a quantum blob is a phase space area equal to $h/2$. Quantum blobs can have arbitrary shapes and sizes and are canonical invariant [15].)

For a pure quantum state for which the wavefunction $\Psi(\boldsymbol{q};t) = \sum_n a_n(t)\Psi_n(\boldsymbol{q})$ is a superposition of quantum states $\Psi_n(\boldsymbol{q})$ the WDF

$$W(\boldsymbol{q},\boldsymbol{p};t) = \sum_{n,m} a_n^*(t)a_m(t)W_{nm}(\boldsymbol{q},\boldsymbol{p}), \tag{7}$$

where

$$W_{nm}(\boldsymbol{q},\boldsymbol{p}) = \frac{1}{h^n}\int \Psi_n^*\left(\boldsymbol{q}-\frac{\boldsymbol{x}}{2}\right)\Psi_m\left(\boldsymbol{q}+\frac{\boldsymbol{x}}{2}\right)\exp(-i\boldsymbol{p}\boldsymbol{x}/\hbar)d\boldsymbol{x}, \tag{8}$$

can be decomposed in a sum of auto-terms $W_{auto}(\boldsymbol{q},\boldsymbol{p};t) = \sum_n |a_n(t)|^2 W_n(\boldsymbol{q},\boldsymbol{p})$ that gathers all terms in (7) with $n = m$ and that corresponds to the sum of the individual WDFs $W_n(\boldsymbol{q},\boldsymbol{p})$ of the interfering states $\Psi_n(\boldsymbol{q})$, and the remaining interference terms; the interference terms

vanish for a mixed quantum state. In particular, when $\Psi_n(q)$ are orthonormal eigenstates of the quantum system the WDF forms also a complete, orthonormal set, in the sense that

$$\int W_{nm}(q,p) W^*_{n'm'}(q,p) dq dp = h^{-n} \delta_{nn'} \delta_{mm'}, \tag{9a}$$

$$\sum_{n,m} W_{nm}(q,p) W^*_{nm}(q',p') = h^{-n} \delta(q-q') \delta(p-p'). \tag{9b}$$

It is important to note that interference between two quantum states along $q$ or $p$ occurs if the individual non-overlapping WDFs of the two states have common projections along $q$ or $p$, respectively [7,16], while transitions between quantum states occur only when their individual WDFs overlap [17]; this distinction between interference and transition, most clearly emphasized in phase space, is illustrated in Fig.1 for $n = 1$.

**3. Phase space formulation of the quantum filtering process**

A filtering device is any device that influences the evolution of an incident quantum state such that the outgoing wavefunction has some "memory" of its original form. A filter actively manipulates the result of a subsequent measurement since it alters the quantum wavefunction.

The formulation in phase space of the action of a filtering device depends on how it affects the incident quantum wavefunction. More precisely, if the incident quantum wavefunction in the position representation, $\Psi_{in}(q)$, is altered by a filter with a transmission function $\Psi_f(q)$, the result of the filtering process, which generates an output wavefunction $\Psi_{out}(q) = \Psi_{in}(q) \Psi_f(q)$, can be described in phase space as

$$W_{out}(q,p) = \int W_{in}(q,p') W_f(q,p-p') dp'. \tag{10}$$

Here $W_{in}$, $W_f$, $W_{out}$ are the WDFs of the incident quantum state, filtering device, and outgoing quantum state, respectively. Note that filtering is in this case described in phase space as a mere multiplication along the $q$ and as a convolution along the momentum direction.

Similarly, the transformation performed by a filter with a transmission function $\Phi_f(\boldsymbol{p})$ in the momentum space on an incident quantum wavefunction $\Phi_{in}(\boldsymbol{p})$ in the momentum representation, i.e. $\Phi_{out}(\boldsymbol{p}) = \Phi_{in}(\boldsymbol{p})\Phi_f(\boldsymbol{p})$, can be represented in phase space as

$$W_{out}(\boldsymbol{q},\boldsymbol{p}) = \int W_{in}(\boldsymbol{q}',\boldsymbol{p})W_f(\boldsymbol{q}-\boldsymbol{q}',\boldsymbol{p})d\boldsymbol{q}'. \tag{11}$$

The convolution is now carried out along the position direction in phase space, the transformation along $\boldsymbol{p}$ being a simple multiplication.

More general filtering processes, which generate output wavefunctions of the form

$$\Psi_{out}(\boldsymbol{q}) = h^{-n/2}\int \Psi_{in}(\boldsymbol{q}')\Psi_f(\boldsymbol{q}-\boldsymbol{q}')\exp(i\boldsymbol{p}_0\boldsymbol{q}'/\hbar)d\boldsymbol{q}' \tag{12}$$

or

$$\Phi_{out}(\boldsymbol{p}) = h^{-n/2}\int \Phi_{in}(\boldsymbol{p}')\Phi_f(\boldsymbol{p}-\boldsymbol{p}')\exp(-i\boldsymbol{q}_0\boldsymbol{p}'/\hbar)d\boldsymbol{p}'. \tag{13}$$

can be represented in phase space as

$$W_{out}(\boldsymbol{q},\boldsymbol{p}) = \int W_{in}(\boldsymbol{q}',\boldsymbol{p}-\boldsymbol{p}_0)W_f(\boldsymbol{q}-\boldsymbol{q}',\boldsymbol{p})d\boldsymbol{q}' \tag{14}$$

or

$$W_{out}(\boldsymbol{q},\boldsymbol{p}) = \int W_{in}(\boldsymbol{q}-\boldsymbol{q}_0,\boldsymbol{p}')W_f(\boldsymbol{q},\boldsymbol{p}-\boldsymbol{p}')d\boldsymbol{p}', \tag{15}$$

respectively.

Note that the phase space representation of the filtering process is different from interference since the WDFs of the incident quantum state and the filtering device must (at least partially) overlap. It differs also from transition since, unlike filtering, quantum transition has no "memory" of the original wavefunction except for the transition probability.

We have assumed that filtering can be represented as the action of a transmission function in position or momentum representations. A more sophisticated description of the filtering process is not necessary as long as the external motion of quantum states is considered. On the other hand, appropriately designed set-ups can be employed to discern between various internal quantum states that become separated in either the position or momentum space. For example, a Stern-Gerlach experimental set-up deflects different spin component values along different directions and thus separates them spatially. Therefore, although internal quantum variables have not been yet introduced in the WDF definition, the phase space formulation of the filtering process can be used to describe also (the majority of) experimental results that involve their direct observation.

As mentioned previously, a filtered wavefunction retains some information of its original state. This is not what happens during a detection process; the result of the measurement is in this case the squared modulus of the quantum wavefunction that is in general a filtered version of the incident wavefunction. More precisely, the positive definite function

$$\int W_{in}(\boldsymbol{q'},\boldsymbol{p'}) W_d(\boldsymbol{q}-\boldsymbol{q'},\boldsymbol{p}-\boldsymbol{p'}) d\boldsymbol{q'} d\boldsymbol{p'} = h^{-n} |\int \Psi_{in}(\boldsymbol{q'}) \Psi_d^*(\boldsymbol{q}-\boldsymbol{q'}) \exp(-i\boldsymbol{p}\boldsymbol{q'}/\hbar) d\boldsymbol{q'}|^2, \qquad (16)$$

where $\Psi_d(\boldsymbol{q})$ and $W_d(\boldsymbol{q},\boldsymbol{p})$ are the wavefunction of the detector in the position representation and its WDF at the moment of detection, can be regarded as a general description of the measurement result of the input quantum state with a detector that inherently filters it in both position and momentum coordinates. This phase space formulation of quantum measurement, which is a convolution of the initial quantum WDF and the WDF of the detector along both coordinate and momentum directions, is known from Refs.18 and 19. It is instructive to compare the expression of the detection process (16) with the probability of transition

$$P_{12} = h^n \int W_1(\boldsymbol{q},\boldsymbol{p})W_2(\boldsymbol{q},\boldsymbol{p})d\boldsymbol{q}d\boldsymbol{p} \tag{17}$$

between an initial and a final state characterized by WDFs $W_1$ and $W_2$, respectively. According to (16) and (17), the detection process is a generalized quantum transition during which the quantum particle "jumps" from the input state to the detector, its presence being observed by a "click" or some other manifestation. However, if the WDF is understood as phase space probability amplitude, the representation of quantum transitions as overlap between the WDF functions of the initial and final states can be viewed as a result of phase space mismatch between different regions. The uncomfortable concept of quantum jump seen as a discontinuity in the particle evolution can be replaced by the interpretation of transition as a change of the WDF due to interaction, for example; the incident quantum particle may or may not follow this change depending if its phase space area of localization (its WDF) fits or not the new area imposed by interaction. This interpretation of quantum transitions should be paralleled with the passage of a guided classical light beam through an interface between waveguides with different spatial/refractive characteristics. The form of the incident light beam is "remembered" only through the excitation coefficients of the guided modes in the new beam.

**4. Phase space description of the double-slit interference experiment**

The double-slit interference experiment reveals either the wave or the particle behavior of a quantum system depending on the number of slits (two or one, respectively) the particles are allowed to pass through. This experiment is fully understood in the realm of standard quantum theory if the wavefunction is regarded as probability amplitude in the spatial domain.

The same experiment can be understood in a phase space formulation of quantum mechanics if the WDF is regarded as probability amplitude in phase space. To demonstrate this affirmation we treat the slits and the detectors as filtering devices in phase space and consider one-dimensional quantum states. More precisely, let us assume that a normalized Gaussian wavefunction $\Psi_{in}(q) = (\pi q_i^2)^{-1/4} \exp(-q^2/2q_i^2)$ with a spatial extent $q_i$ and a WDF $W_{in}(q,p) = (2/h)\exp(-q^2/q_i^2 - p^2 q_i^2/\hbar^2)$ is incident on the plane of the slits. The incident WDF, the squared modulus of the incident wavefunction and its Fourier transform (all normalized to their maximum values) are represented in Fig.2 in the normalized coordinates $Q_i = q/q_i$, $P_i = pq_i/\hbar$. Since the two slits do not introduce (ideally) any dephasing, we can model their wavefunction as a superposition of Gaussian transmittances $\Psi_-(q)$ and $\Psi_+(q)$ separated by a distance $2d$, i.e. as a one-dimensional cat-like state

$$\Psi_f(q) = \Psi_+(q) + \Psi_-(q) = N\left[\exp\left(-\frac{(q-d)^2}{2q_f^2}\right) + \exp\left(-\frac{(q+d)^2}{2q_f^2}\right)\right], \tag{18}$$

where $N = (4\pi q_f^2)^{-1/4}[1 + \exp(-d^2/q_f^2)]^{-1/2}$ is the normalization constant. The WDF of (18), given by

$$W_f(q,p) = \frac{\exp(-p^2 q_f^2/\hbar^2)}{h[1+\exp(-d^2/q_f^2)]}\left[\exp\left(-\frac{(q-d)^2}{q_f^2}\right) + \exp\left(-\frac{(q+d)^2}{q_f^2}\right) + 2\exp\left(-\frac{q^2}{q_f^2}\right)\cos\left(\frac{2dp}{\hbar}\right)\right]$$

$$= W_+(q,p) + W_-(q,p) + W_{int}(q,p), \tag{19}$$

is represented in Fig.3 (together with the squared modulus of the wavefunction and its Fourier transform) for $d = 4q_f$ in adimensional coordinates $Q_f = q/q_f$, $P_f = pq_f/\hbar$. It consists (for a sufficiently large separation $d$) of two outer terms, $W_-(q,p)$ and $W_+(q,p)$, which represent the WDFs of the individual slits, and of an interference term $W_{int}(q,p)$ (the last term in (19)), which suggest the possibility of interference.

If the incident wavefunction is not perfectly centered with respect to the two slits, i.e. it is spatially misaligned with $\Delta$, the WDF of the quantum state after passing through the slits is given by (10):

$$W_{out}(q,p) = K \exp\left(-\frac{(q-\Delta)^2}{q_i^2}\right) \exp\left(-\frac{p^2}{\hbar^2}\frac{q_f^2 q_i^2}{q_f^2+q_i^2}\right)$$

$$\times \left[\exp\left(-\frac{(q-d)^2}{q_f^2}\right) + \exp\left(-\frac{(q+d)^2}{q_f^2}\right) + 2\exp\left(-\frac{q^2}{q_f^2}\right)\exp\left(-\frac{d^2}{q_f^2+q_i^2}\right)\cos\left(\frac{2dp}{\hbar}\frac{q_i^2}{q_f^2+q_i^2}\right)\right]$$

(20)

with $K = h^{-1}[1+\exp(-d^2/q_f^2)]^{-1}[\pi(q_i^2+q_f^2)]^{-1/2}$. From (20) it follows that $W_{out}(q,p)$ is a scaled version of the WDF of the slits for $d, q_f \ll q_i$, i.e. for the case when the incident quantum state passes through both slits; this is the case for all double-slit interference experiments. So, we can safely consider that the WDF of the quantum wavefunction after the slits is given by equation (19) ($W_{out}(q,p)$ is identical to $W_f(q,p)$ apart from a proportionality factor that can be ignored) and is represented in Fig.3. The two individual Gaussian states can interfere due to the existence of the interference term in the WDF, which has larger amplitude than the individual WDFs of the two slits ($W_-(q,p)$ and $W_+(q,p)$) and attains its maximum value at the phase space origin $q=0$, $p=0$. However, interference immediately after the slits can only occur in the momentum space, since $W_-(q,p)$ and $W_+(q,p)$ have common projections only along $p$. In an alternative description, as shown in Fig.3, oscillatory behavior of the squared modulus of the outgoing quantum wavefunction is observed only in the momentum space and not in the coordinate space. To obtain interference along $q$ the two spatially separated beams must overlap, i.e. their WDFs must have also common projections along $q$. This is achieved after free space propagation for a distance $D$ along the $z$ direction, the corresponding transformation law for the WDF between the plane of the slits located at $z=0$ and the plane of interference located at $z=D$ being given by $W(Q_f, P_f; z=D) = W(Q_f - DP_f/q_f, P_f; z=0)$ in the normalized coordinates $Q_f$, $P_f$. This is a shear transform of the WDF along $q$, interference in the spatial coordinate appearing as soon as the WDFs of the individual slits $W_-(q,p)$ and $W_+(q,p)$ begin to overlap along $q$. Fig.4 shows the WDF of the quantum state at the plane $z = D = 5q_f$ after the slits and the squared modulus of the quantum wavefunction in both coordinate and momentum representations. It is clear that interference appears in the spatial coordinate. (Note that free space propagation does not alter the squared modulus of $\Phi(p)$.) If a photographic plate is placed at this distance after the slits an interference pattern will be observed.

What happens when a detector is placed at a smaller distance from the plane of the slits, before the individual WDFs begin to overlap? The action of a detector (particle or photon counter) placed, say, at $z = 0$ in front of one slit can be modeled as follows: it allows the passage of particles/photons from the other slit and prevents further propagation of particles/photons that pass through the slit in front of which it is placed. A Gaussian transmittance centered on the slit after which no detector is placed, with a spatial extent larger than that of the slit, is a convenient model of such a detector; the detection process is again considered as a phase space filtering process due to a transmittance with a spatially misaligned (with respect to the slits) normalized wavefunction $\Psi_d(q) = (\pi q_d^2)^{-1/4} \exp[-(q-d)^2/2q_d^2]$. (This is possible since, unlike the phase space treatment of the detection process expressed by (16), the measuring device detects only part of the total wavefunction and thus acts as a filter for the total wavefunction.) Fig.5 shows the transmittance of a detector with $q_d = 2.4q_f$ (the transmittance of the slits is represented with dotted line for comparison); this detector allows only the passage of the particles that pass through a single slit. The result of filtering the WDF in (19) with a Gaussian filter is again given by an expression of the same form as (20) with $\Delta$ replaced by $d$ and $q_i$ replaced by $q_d$; the filtered wavefunction, however, differs this time substantially from the WDF of the two slits since $q_d \cong d$. The result of simulations for $q_d = 2.4q_f$ is shown in Fig.6 together with the squared modulus of the filtered wavefunction and of its Fourier transform. No interference is present in either position or momentum space and no interference is expected to appear after free space propagation of any length since the filtered WDF consists practically only from $W_+(q,p)$. It is important to note that not only the $W_-(q,p)$ contribution to the WDF of the slits is filtered away but so is also the interference term although the transmission function is different from zero on the phase space region occupied by $W_{int}(q,p)$. Filtering devices, as emphasized also in Ref.7, do not act directly on interference terms; phase space regions on which quantum particles/photons are localized (in our case the phase space regions corresponding to $W_+(q,p)$ and $W_-(q,p)$) are filtered away, the interference disappearing because the remaining quantum state has no other state to interfere with. (Note that the phase space region occupied by the interference term does not correspond to any significant probability of finding the quantum particles that pass through one or the other slit.) A detector placed in front of the remaining quantum state at any distance from the plane of the slits does not record any interference pattern since $W_{int}(q,p)$ has disappeared.

Thus, the phase space formulation of quantum interference shows in a more intuitive form than standard quantum mechanics that any attempt to detect the slit the particles have passed through results not only in blocking the further passage of particles that pass through the slit after which the detector is placed but also in the disappearance of any possibility of interference. The interference term in WDF indicates the possible occurrence of interference; it actually takes place along $q$ and/or $p$ only when the interfering states have common projection along the corresponding coordinate(s).

If the detector does not prevent the further passage of all particles that hits it, situation modeled by a larger $q_d$ (see Fig.7 with $q_d = 4q_f$) the interference term in the WDF is revived, even if only a small fraction of particles pass through the detector. Fig.8 simulates this case; this situation corresponds to the occurrence of interference fringes (at a distance from the slits where the two beams are superimposed) with visibility less than 1 and is perhaps the nearest practical situation to the wave-particle duality concept.

**5. Phase space representation of delayed-choice experiments**

To be specific we consider the delayed-choice experiment sketched in Fig.9 [20]. The incident one-dimensional wavefunction passes through the two slits followed by identical off-axis lenses that bring the interfering beams in superposition at a certain distance $z = D$ after the plane of the slits (the two lenses can be eventually replaced by a single large lens centered with respect to the two slits). A quantum particle manifests itself as a wave, i.e. quantum interference is observed, if a photographic plate is introduced in the region of superposition, or manifest itself as a particle if its position at a plane $z > D$, where the two beams no longer overlap (their WDF have no longer common projections along $q$), is detected by a particle counter. The wave-particle duality is emphasized by the fact that the choice of introducing one measurement device or another can be made after the particle passes through the slits. According to the phase space interpretation, however, this delayed choice does not influence the nature of the quantum particle but only establishes the distance from the slits at which the detection is performed (and hence establishes the fact that the detection is made when the WDF of the two wavefunctions have or not common projections along $q$, i.e. when two wavefunctions interfere or not).

Assuming that the lenses are situated immediately after the slits, the quantum wavefunction immediately after the lenses can be approximated with

$$\Psi_{out}(q) \approx \exp(-iq^2/K^2)\{\exp[-(q-d)^2/q_f^2 - ip_0 q] + \exp[-(q+d)^2/q_f^2 + ip_0 q]\} \qquad (21)$$

where $K$ characterizes the refractive power of the lenses and $p_0$ is a measure of the off-axis displacement.

Figs.10-12 show the WDF and the squared modulus of the wavefunction and its Fourier transform immediately after the lenses, at $z = D = q_f$, where the partial wavefunctions that pass through different slits are brought into superposition, and at $z = 3q_f$ where they no longer interfere in the spatial domain. All simulations are made for $K = 2q_f$, $p_0 = 3\hbar/q_f$. In this case the presence of the off-axis lenses prevents interference along p immediately after the slits as well as after any distance of free space propagation. The interpretation of all these situations parallels the discussions in the previous section: immediately after the slits interference does not occur along *q* or along *p*, but it is possible due to the interference term. After free space propagation along a distance equal in this case to $q_f$ the two beams are in superposition and interference in the spatial domain can be observed with a properly placed photographic plate. At larger distances the two beams become again spatially separated and it is possible to detect the particles in only one beam by a particle counter. In this case, as discussed previously, no interference will be observed if the detector blocks all particles, interference being possible otherwise if the two beams are subsequently superimposed. The main advantage of the phase space interpretation of the delayed-choice experiment is, however, that the nature of the particle is not questioned. The form of the WDF, which is the probability amplitude in phase space, dictates the outcome of the measurement. One can safely decide if a photographic plate should or not be placed at $z = D$ as long as the quantum wavefunction does not reach this plane, because the interaction of the quantum wavefunction with any filtering device that alters its propagation takes place continuously, when any filter is encountered, and not only at the detector. The evolution of a quantum wavefunction through a set-up consisting from several filters is analogous to passing a classical light beam through several masks.

# 6. Discussions and conclusions

We have presented a phase space treatment of the filtering process upon a quantum wavefunction, which offers an interesting insight in the wave-particle duality concept. This treatment has revealed that there is no duality, i.e. the quantum particle does not manifest itself as a "wave" or as a "particle" depending on the experimental conditions. Rather its WDF, which represents the phase space probability amplitude, has or not an interference term depending on the filtering devices that are introduced in the experimental set-up. The quantum particle, through its wavefunction or its WDF, interacts with these filtering devices during its entire evolution, from its generation to the final detection process; the "wave" or "particle" character does not suddenly appear at detection but is influenced by any filtering that affects the wavefunction throughout the set-up. The results of both quantum interference and delayed-choice experiments can be fully explained in the phase space formulation of quantum mechanics by endowing the WDF with the significance of phase space probability amplitude. Although all simulations of the slits and detector have been performed with Gaussian wavefunctions for simplicity (analytical results can be obtained easily in this case) the use of more adequate descriptions of filtering devices should not alter the conclusions of this paper. In the examples only quantum wavefunctions in the position representation have been chosen, but as evident from Section 3, the momentum representation is equally welcome. Note that a beam splitter can be regarded as an analog in the momentum space of two slits in the position space.

Although we have assumed throughout the paper that the filtering device, in particular the slits and the particle counter, is quantum, this is not necessarily true. Since the variables of the quantum and classical phase spaces are the same a classical measuring apparatus can be formally treated in the same way as a quantum one; in particular all the results (the expressions (10), (11), (14), (15) and (16)) remain valid as long as the classical measuring apparatus can be characterized by a WDF. This might turn out to be of considerable advantage since it is quite difficult to calculate from quantum considerations the quantum wavefunction or the WDF of sizeable devices; a classical description could be more appropriate for the determination of the WDF in these cases and hence the accommodation of classical filtering devices in a quantum theory of filtering is invaluable. Note that phase space descriptions of both classical light beams [21] and classical ensembles of particles [21] are already well developed. For classical states containing a large number of quantum particles the WDF should, however, no longer be regarded as phase space probability amplitude but as the phase space representation of the object. Then, negative values of the WDF could exist (and they are known and experimentally proven in classical optics [21]) or not depending on the existence or absence of correlations between adjacent quantum blobs (see the interpretation of negative WDF values in [14]).

Since a macroscopic state is generally formed from a large number of quantum constituents it follows that the classical WDFs, as well as the quantum WDFs, must be restricted to a phase space area not smaller than a quantum blob, which implies that the preparation and/or detection of an eigenstate of $\hat{q}$ or $\hat{p}$ (or of operators that depend linearly on position or momentum) is not possible. Such an eigenstate as well as the corresponding WDF would be a $\delta$-function in the $q$ or $p$ domain (which has no phase space area), situation forbidden by Heisenberg's uncertainty that sets a minimum value for the phase space area of quantum (or classical) states. This consequence seems to be odd for a classical filtering or measuring device because quantum effects are not apparent in the classical realm. However, in phase space the formal distinction between quantum and classical systems disappears and the fact that quantum effects are not observable for sizeable systems can be explained by the fact that the minimum phase space area imposed by the Heisenberg's uncertainty principle is much smaller than the phase space area occupied by a classical device. It follows that measurements/filtering of quantum systems, performed by either quantum or classical devices are not ideal (cannot be represented by $\delta$-functions in phase space) and do influence the quantum wavefunction, as demonstrated by (10), (11), (14), (15) and (16).

**Figure captions**

Fig.1 (a) Two states with individual WDFs $W_1$ and $W_2$ that have common projections along the $q$ and $p$ axes interfere along both spatial and angular coordinates. (b) Transitions are expected to occur if the individual WDFs overlap at least partially.

Fig.2 The WDF, squared modulus of an incident Gaussian wavefunction and its Fourier transform

Fig.3 Same, for a filter that consists from two slits

Fig.4 Same, for the filtered wavefunction at a distance $D = 5q_f$ from the plane of the slits

Fig.5 The transmittance of a detector with width $q_d = 2.4q_f$ (solid line) and of the two slits (dotted line)

Fig.6 The WDF, squared modulus of the wavefunction and its Fourier transform for the state filtered by the detector in Fig.5

Fig.7 Same as in Fig.5, but for $q_d = 4q_f$

Fig.8 The WDF, squared modulus of the wavefunction and its Fourier transform for the state filtered by the detector in Fig.7

Fig.9 Set-up for the delayed-choice experiment

Fig.10 The WDF, squared modulus of the wavefunction and its Fourier transform for the quantum state immediately after the lenses in Fig.9

Fig.11 Same as in Fig.10, but after a propagation distance $z = q_f$ from the plane of the slits

Fig.12 Same as in Fig.11, for $z = 3q_f$

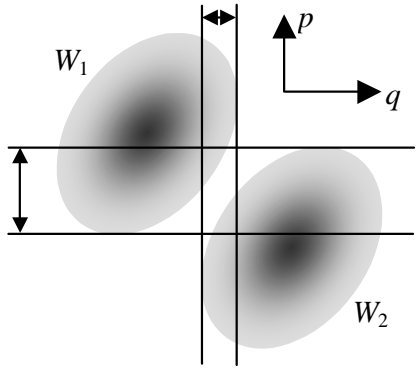 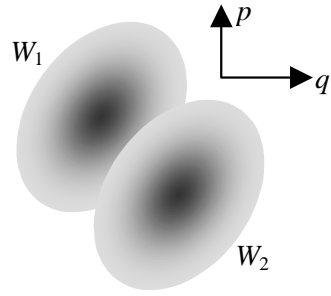

(a)                                             (b)

Figure 1

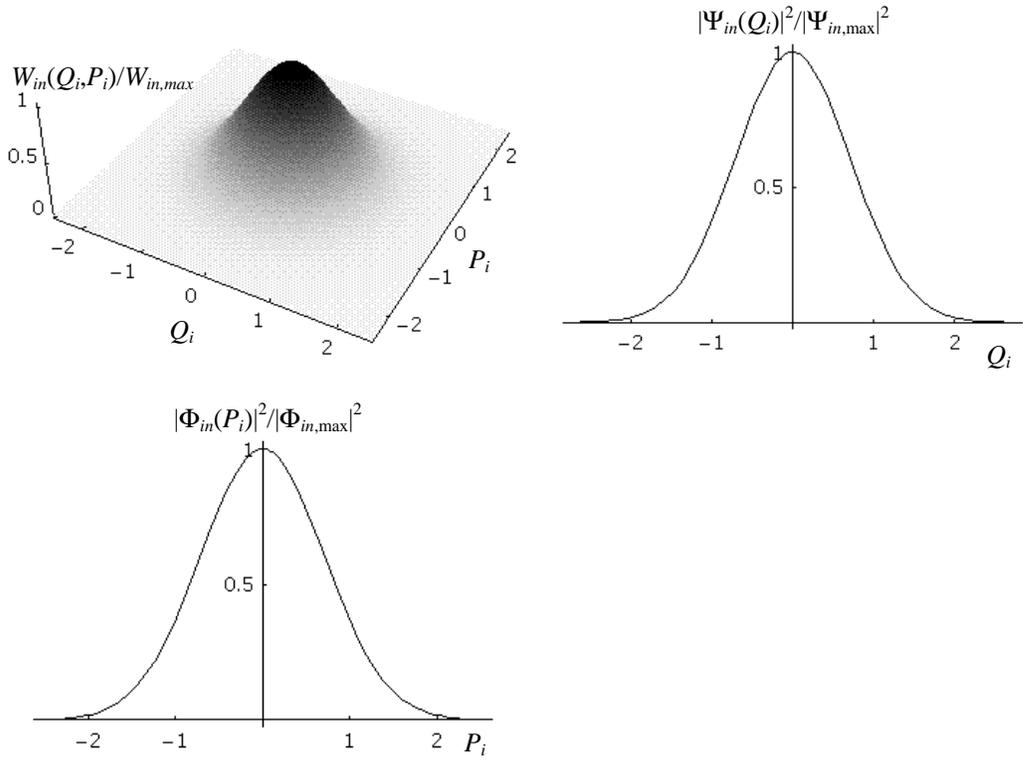

Figure 2

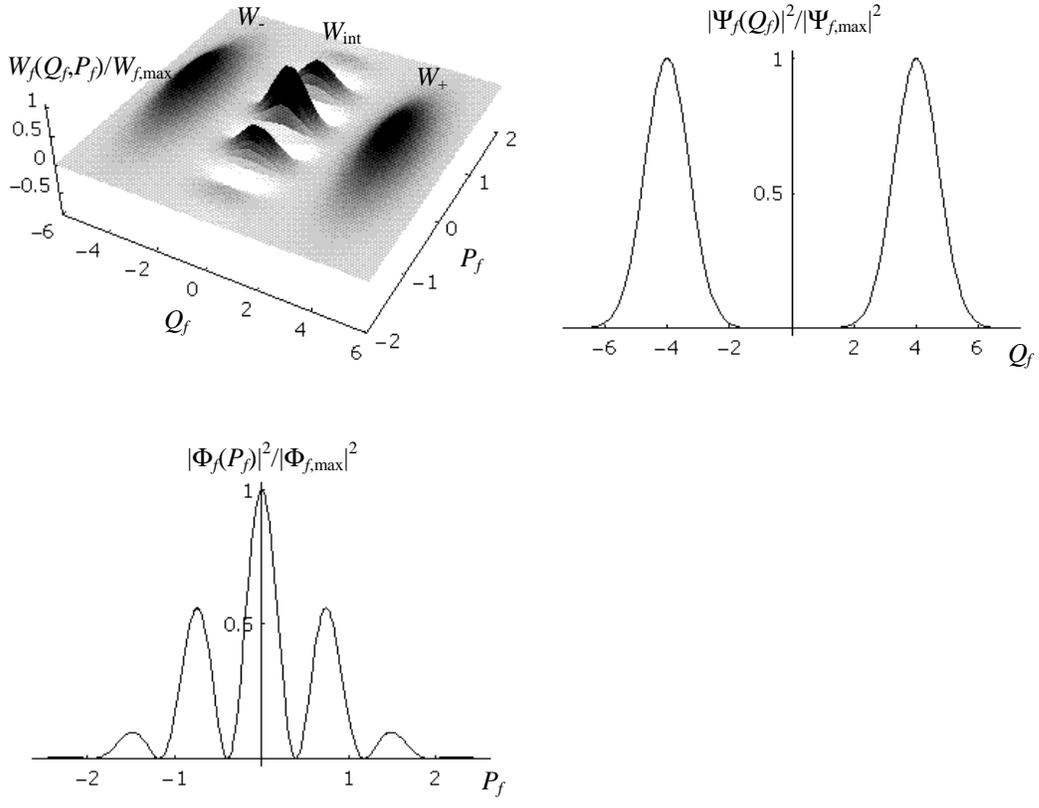

Figure 3

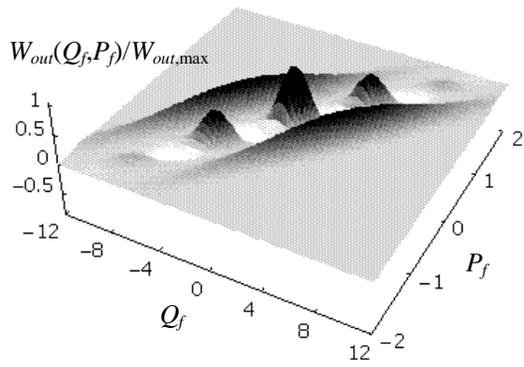
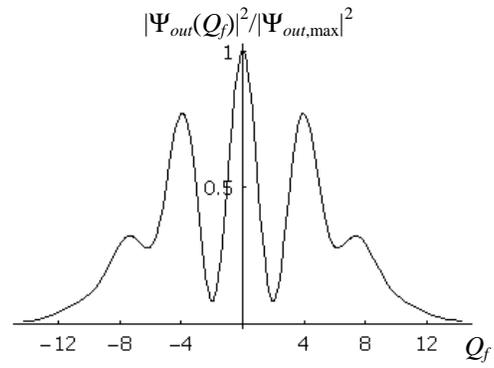
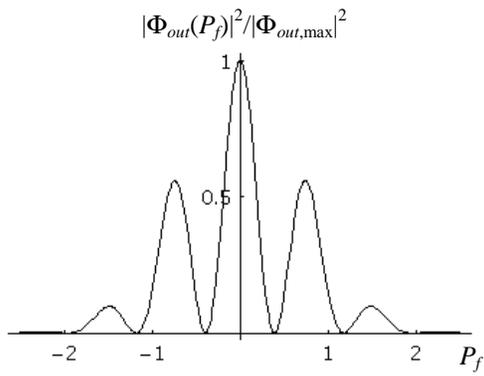

Figure 4

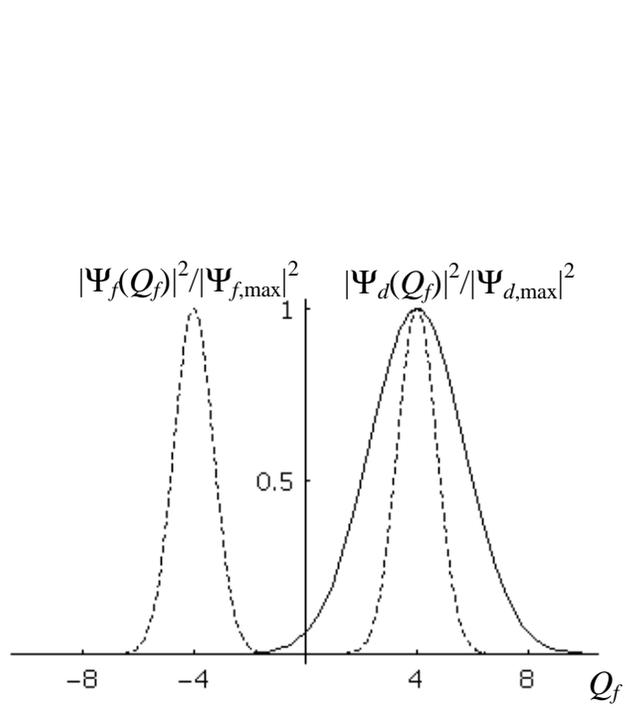

Figure 5

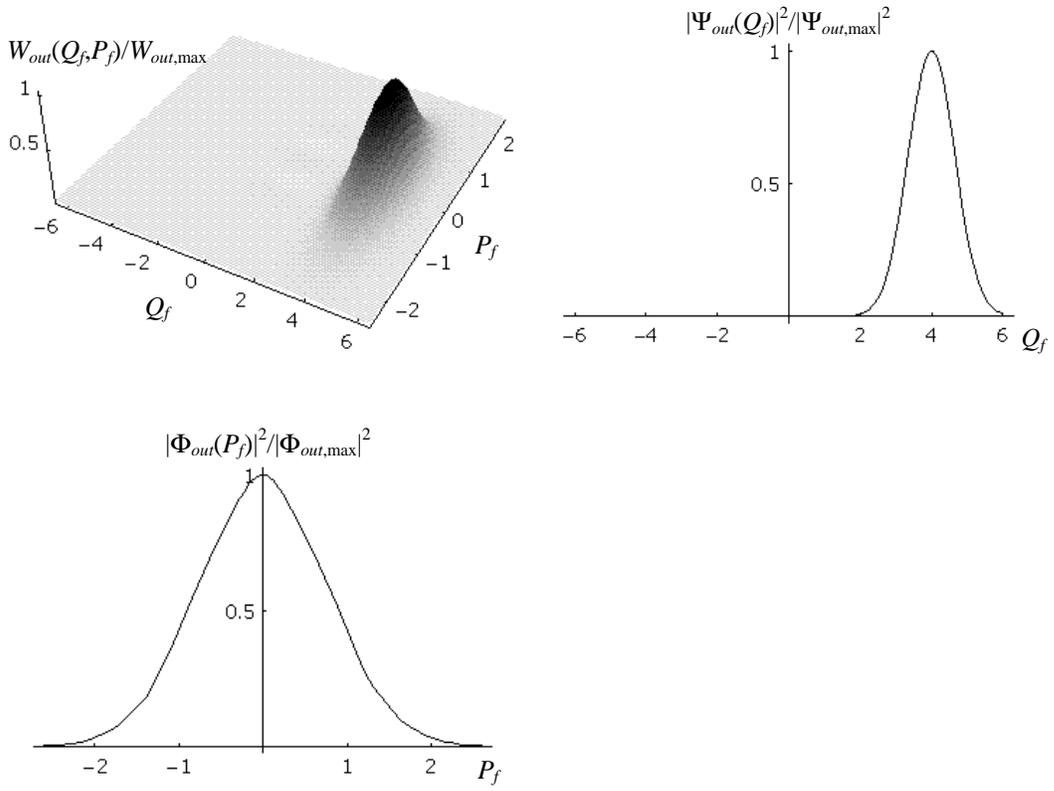

Figure 6

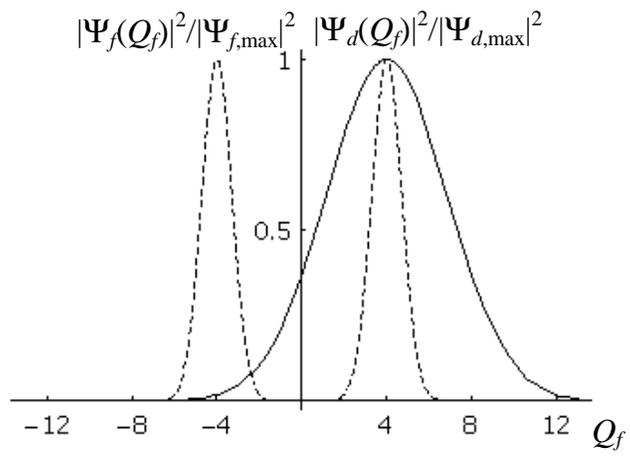

Figure 7

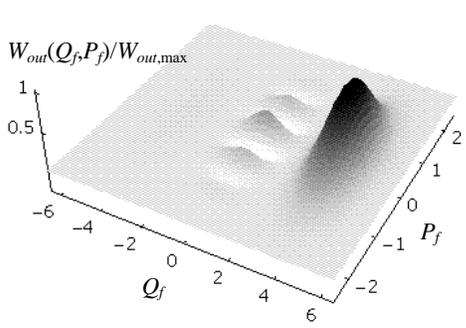
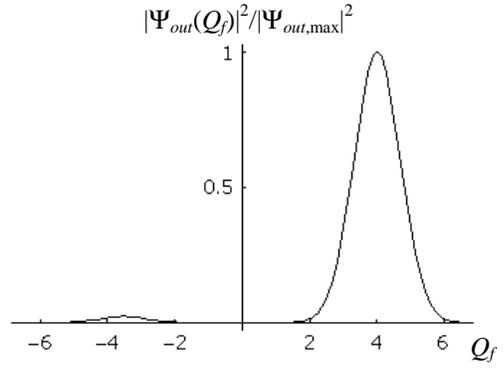
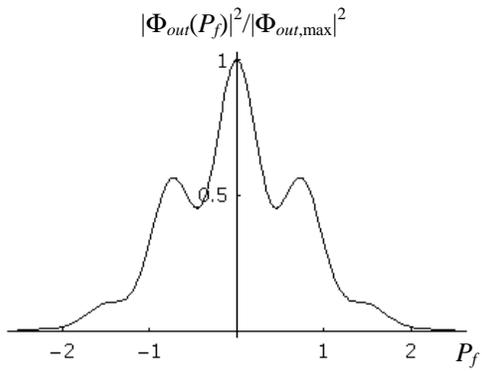

Figure 8

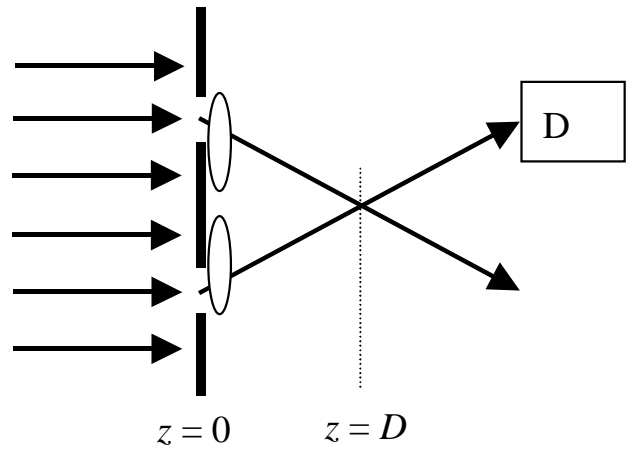

Figure 9

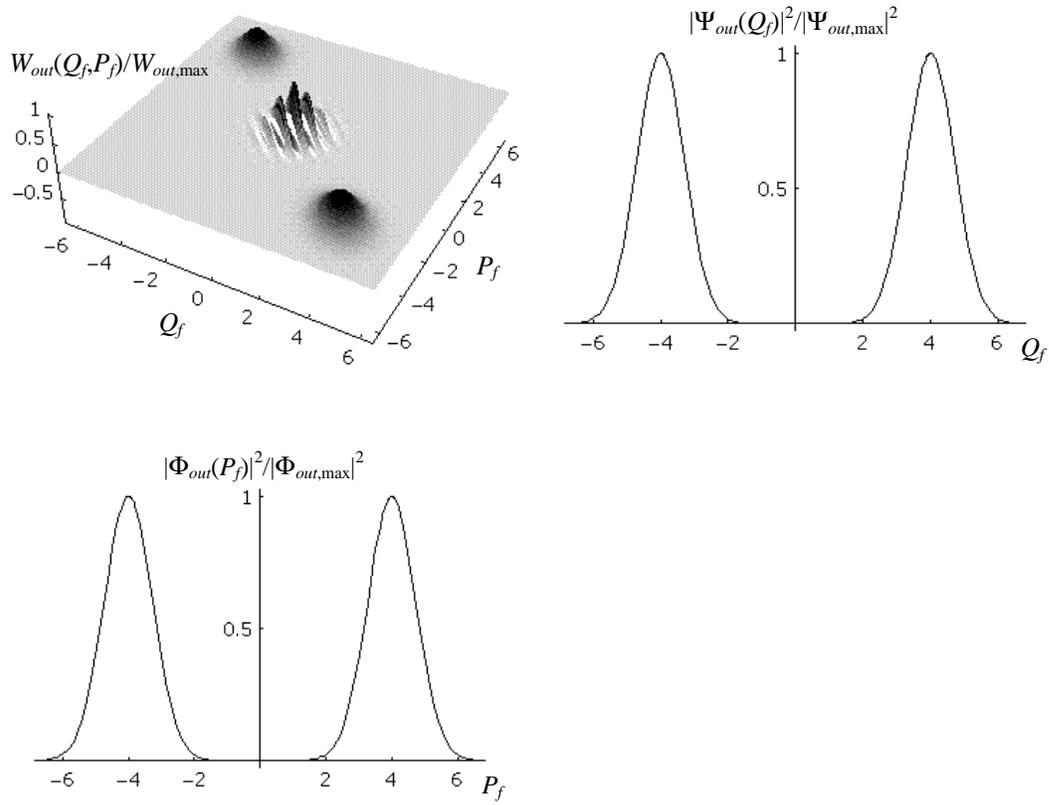

Figure 10

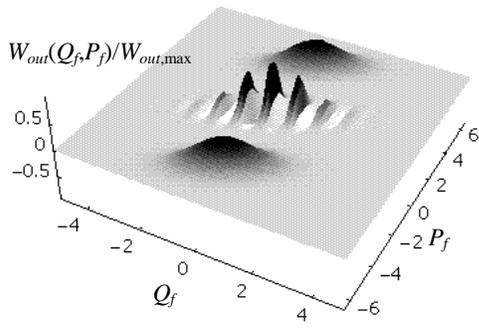
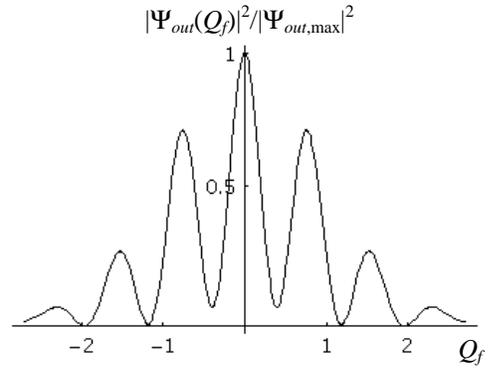
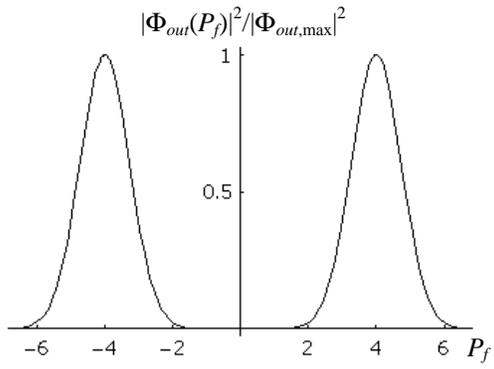

Figure 11

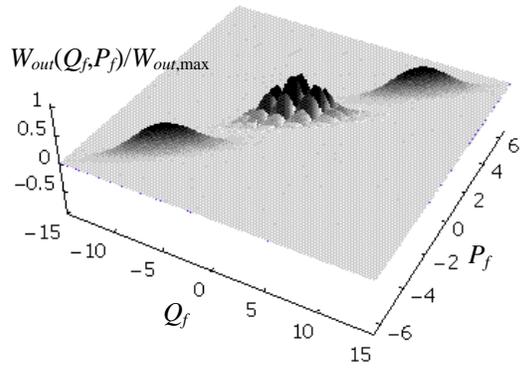
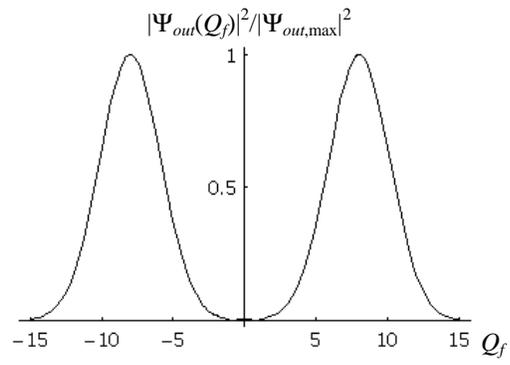
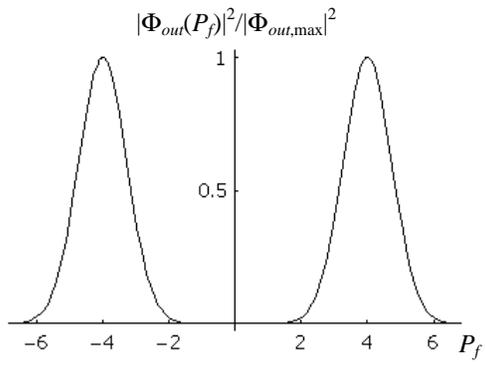

Figure 12